\documentstyle[12pt,aas2pp4]{article}
\def\kms{\relax \ifmmode {\,\rm km\,s}^{-1}\else \,km\,s$^{-1}$\fi}

\def\farcs{\hbox{$.\!\!^{\prime\prime}$}}

\def\secd#1.#2{ #1\farcs#2 } 

\def\mincir{\ \raise-2.truept\hbox{\rlap{\hbox{$\sim$}}\raise5.truept
    \hbox{$<$}\ }}
\def\magcir{\ \raise-2.truept\hbox{\rlap{\hbox{$\sim$}}\raise5.truept
    \hbox{$>$}\ }}
\def\gr{$^\circ$}

\def\nii{[N {\sc ii}]}
\def\sii{[S {\sc ii}]}

\def\heii{He {\sc ii}}
\def\oiii{[O {\sc iii}]}
\def\ha{H$\alpha$}

\lefthead{NGC 6337, He 2-186, K 4-47}
\righthead{Corradi et al.}

\tightenlines

\begin{document}

\title{High-velocity collimated outflows in planetary nebulae:\\ NGC~6337,
He~2-186, and K 4-47\footnote{Based on observations obtained at the 3.5-m New
Technology Telescope (NTT) of the European Southern Observatory, and the at
the 2.6-m Nordic Optical Telescope (NOT) operated on the island of La Palma
by NOTSA, in the Spanish Observatorio del Roque de Los Muchachos of the
Instituto de Astrof\'\i sica de Canarias, and with the NASA/ESA {\it Hubble
Space Telescope}, obtained at the Space Telescope Science Institute, which is
operated by AURA for NASA under contract NAS5-26555.}}

\author{Romano L.M. Corradi, Denise R. Gon\c calves, Eva Villaver, Antonio
Mampaso}

\affil{Instituto de Astrof\'{\i}sica de Canarias, c. Via Lactea S/N, \\
	E--38200 La Laguna, Tenerife, Spain. \\e--mail: rcorradi@ll.iac.es,
	denise@ll.iac.es, villaver@ll.iac.es, amr@ll.iac.es}

\author{Mario Perinotto}

\affil{Dipartimento di Astronomia e Scienza dello Spazio, Universit\`a 
di Firenze, \\Largo E. Fermi 5, 50125 Firenze, Italy. 
\\e--mail: mariop@arcetri.astro.it}

\author{Hugo E. Schwarz}

\affil{Nordic Optical Telescope, Apartado 474, E--38700 Sta. Cruz de La 
             Palma, Spain\\e--mail: hschwarz@not.iac.es}

\author{Caterina Zanin}

\affil{Institut f\"ur Astronomie, Technikerstr. 25, A-6020 Innsbruck, Austria
\\e--mail: zanin@ast1.uibk.ac.at}


\begin{abstract}
We have obtained narrow-band images and high-resolution spectra of the planetary
nebulae NGC~6337, He~2-186, and K~4-47, with the aim of investigating the
relation between their main morphological components and several
low-ionization features present in these nebulae.

The data suggest that NGC~6337 is a bipolar PN seen almost pole on, with polar
velocities of $\ge$ 200~\kms. The bright inner ring of the nebula is
interpreted to be the ``equatorial'' density enhancement. It contains a
number of low-ionization knots and outward tails that we ascribe to dynamical
instabilities leading to fragmentation of the ring or transient density 
enhancements due to the interaction of the ionization front with  previous 
density fluctuations in the ISM.  The lobes show a
pronounced point-symmetric morphology and two peculiar low-ionization
filaments whose nature remains unclear.

The most notable characteristic of He~2-186 is the presence of two
high-velocity ($\ge$$135$~\kms) knots from which an S-shaped lane of emission
departs toward the central star.

K~4-47 is composed of a compact core and two high-velocity, low-ionization 
 blobs. We interpret the substantial broadening of line
emission from the blobs as a signature of bow shocks, and using the modeling
 of Hartigan, Raymond, \& Hartman (1987), we derive a shock velocity of 
 $\sim$150~\kms\ and a
mild inclination of the outflow on the plane of the sky.

We discuss possible scenarios for the formation of these nebulae and their
low-ionization features. In particular, the morphology of K~4-47 hardly fits
into any of the usually adopted mass-loss geometries for single AGB stars.  Finally,
we discuss the possibility that point-symmetric morphologies in the lobes
of NGC~6337 and the knots of He~2-186 are the result of precessing outflows
from the central stars.

\end{abstract}

\keywords{planetary nebulae: individual (NGC 6337, He 2-186, K 4-47) - 
ISM: kinematics and dynamics - ISM: jets and outflows}

\section{Introduction}

Ionized gas in planetary nebulae (PNe) expands with typical velocities
between 10 and 40~\kms\ (cf. Weinberger 1989; Corradi \& Schwarz 1995),
values which are similar or only slightly larger than those of the winds
from their AGB progenitors, indicating moderate acceleration of the gas
during the PN phase.  Some PNe, however, exhibit expansion velocities an
order of magnitude larger (100 -- 500 \kms) and are generally associated with
collimated outflows, such as bipolar PNe (cf. Corradi \& Schwarz 1995) or
jets (e.g., Bryce et al. 1997; Schwarz et al. 1997). These high-velocity,
collimated outflows are usually best observed in the light of low-ionization
species such as the \nii\ 658.3~nm line. The dynamical mechanisms collimating
and accelerating the gas to the high velocities observed are not yet clear
(Garc\'\i a-Segura et al. 1999, Soker 1997).  

In this paper, we present a morphological and kinematical study of three PNe
(NGC\-~6337, He~2-186, and K~4-47) in which we also found the occurrence of
bipolar outflows expanding with notably high velocities. This work is part of
an observational program aimed at studying  small-scale, low-ionization
structures in PNe, such as knots, bullets, filaments, ansae, etc.  These
structures are very puzzling morphological components of
PNe (Balick et al. 1998, and references therein). Most of the targets of our
program were selected by Corradi et al. (1996b) by computing
(\nii+\ha)/\oiii\ ratio maps in the image catalogs of Schwarz, Corradi, \&
Melnick (1992) and Gorny et al. (1999).  Results for four PNe were reported
in two previous papers (Corradi et al. 1997; Corradi et
al. 1999). In brief, some of the low-ionization features appear to be 
associated with extended jets (NGC 3918, and K 1-2) or with multiple
collimated outflows (IC 4593 and Wray 17-1) characterised by moderate to high
expansion velocities.  In this paper, we discuss the properties of the
low-ionization features in NGC~6337, He~2-186 and K~4-47.  

 
\section{Observations}

Images and long-slit spectra of NGC~6337 (PN G349.3-01.1), were obtained on
1996 April 27 at ESO's 3.5-m New Technology Telescope (NTT) at La Silla
(Chile), using the EMMI multimode instrument.  With the TEK 2048$^2$ CCD
ESO\#36, the spatial scale of the instrument was 0$''$.27~pix$^{-1}$ both for
imaging and spectroscopy.  The central wavelength and full width at
half-maximum (FWHM) of the \nii\ filter used for imaging are 658.8~nm and
3.0~nm, and those of the \oiii\ filter 500.7~nm and 5.5~nm. Further details
of the observations are listed in Table~1.  As with spectroscopy, EMMI was
used in the long-slit, high-resolution mode (Corradi, Mampaso, \& Perinotto
1996a), providing a reciprocal dispersion of 0.004~nm~pix$^{-1}$, and a
spectral resolving power of $\lambda$/$\Delta\lambda$=55000 (5.5~\kms) with
the adopted slit width of 1$''$.0. The slit length was of 6~arcmin. The
echelle order selected by using a broad \ha\ filter includes the \heii\ line
at $\lambda$=656.01~nm, \ha\ at $\lambda$=656.28~nm, and the \nii\ doublet at
$\lambda$=654.81 and 658.34~nm.  The slit was positioned through the center
of the nebula at position angles P.A.=$-39$\gr\ and P.A.=$-75$\gr\ (see
Table~1).

He~2-186 (PN G336.3-05.6) was observed at the NTT on 1991 April 26 also with
the EMMI instrument.  Two medium-resolution spectra centered on \ha\ were
obtained with grating \#6, using a 2$\times$2 binning of the CCD so that the
spatial scale was of 0$''$.70 and the spectral reciprocal dispersion 0.08~nm
per binned pixel, corresponding to a resolving power of 4000 (75~\kms). The
slit width was 1$''$.0. The spectral range covered includes \ha, \heii, the
\nii\ doublet, as well as the \sii\ doublet at $\lambda$=671.6~nm and
673.1~nm.  We also retrieved images of He~2-186 from the {\it HST} archive
obtained on 1999, February 7 with the WFPC2 camera (PC CCD, 0$''$.0455
pix$^{-1}$) in the \nii\ F658N narrowband filter (659.0/2.9~nm). Two images
of 400 and 200~sec exposure time, were combined to improve the S/N ratio and
remove cosmic rays.

K 4-47 (PN G149.0+04.4) was observed at the 2.6-m Nordic Optical Telescope
(NOT) of the Observatorio del Roque de los Muchachos (La Palma,
Spain). Images were taken on 1997 September 20 and October 21, using the
ALFOSC instrument and a Loral 2k $\times$ 2k CCD, providing a scale of
0$''$.19 per detector pixel. The central wavelength and FWHM of the filters
used at the NOT, corrected for the temperature at the time of observations,
were: 500.9/3.0~nm (\oiii), 650.3/2.8~nm (\ha\ off-band), 656.4/0.8~nm (\ha),
658.5/0.9~nm (\nii), and 672.0/5.0~nm (\sii).  Finally, a high resolution
spectrum of K~4-47 was obtained using the echelle spectrograph IACUB (McKeith
et al. 1993) on 1998 January 17.  The detector was a Thompson THX31156 1024
CCD, giving a spatial scale of 0$''$.28 per pixel and a reciprocal dispersion
of 0.005~nm per pixel. The slit width was 0$''$.8, providing a spectral
resolution of 25000 (12~\kms).  Exposure times and slit position angles are
given in Table~1.

Images and spectra were reduced in a standard way using MIDAS and IRAF. 

\section{NGC 6337}

According to the literature, NGC 6337 is a multiple shell planetary nebula
(e.g.,  Stanghellini, Corradi, \& Schwarz 1993).  Its distance, determined by
statistical methods, is in the range of 1.1 to 1.7~kpc (Acker et al. 1992).
Physical parameters for the nebula are also derived
 from statistical studies spread
in the literature (see, for instance, G\'orny, Stasi\'nska, \& Tylenda 1997;
Phillips 1998).  The connection between the central star evolution and the
morphology of NGC~6337 was studied by Stanghellini et al. (1993), who
classified it as an elliptical nebula with multiple shells showing two axes
of symmetry for the external structures.  To our knowledge, no detailed
kinematical studies of NGC~6337 exist.

Our \nii\ and \oiii\ NTT images of NGC~6337 are presented in
Fig.~\ref{F-n6337i}, together with the \nii/\oiii\ ratio map.

\placefigure{F-n6337i}

\subsection{Morphology}

The main morphological component of NGC 6337 is a circular ring. In 
\oiii\ the inner and outer radii of the ring are  14$''$ and 24$''$,
respectively, the surface brightness peaking at 20$''$ from the central star.
In  \nii\  the ring is broken into filamentary structures which are
aligned along the radial direction from the central star. Most of the \nii\
emission is concentrated outside  the \oiii\ peak radius, and shows extended,
outward tails which are neatly enhanced  in the \nii/\oiii\ ratio map.
The most prominent of these low-ionization radial features is labeled as
$C$ in the \nii\ image of Figure~\ref{F-n6337i}.

The ring is surrounded by a faint halo. In  \oiii\  this halo has a
pretty uniform surface brightness and the shape of two ``spiral arms''
bending clockwise toward the NW and SE directions.  In \nii, the halo is more
irregular, and exhibits two peculiar bright filaments, labeled  $A$ and $B$
in Figure~\ref{F-n6337i}. While $B$ is   in rough radial alignment  
with the central star, $A$ is almost perpendicular to the radius from the
central star. The maximum
diameter of the halo is 120$''$.

\subsection{Kinematics}

\placefigure{F-n6337s}

\placefigure{F-n6337C}

The long slit of the spectrograph was positioned through the central star at
P.A.=$-75$\gr, covering the low-ionization feature $C$, and at P.A.=$-39$\gr,
throughout filament $B$ and intersecting a portion of $A$ (see
Fig.~\ref{F-n6337i}). Spectra are shown in Figure~\ref{F-n6337s}. \ha\ and
\nii\ velocities along the slit were measured by Gaussian fitting and are
also presented in Figure~\ref{F-n6337s}.

At both position angles, the most notable feature is the presence of
faint high-velocity  halo components which appear located, in projection, both
inside and outside the bright ring.  These components span a radial velocity
range of $\sim$400~\kms.  The overall shape of the nebula and of the velocity
field is reminiscent of that of a bipolar nebula seen almost pole on, the
bright ring being the projection of the equatorial density enhancement, which
possibly collimates the high velocity outflow (cf. the mildly inclined PN
Hb~5 (Corradi \& Schwarz 1993b) and with the pole on bipolar PN K 4-55
(Guerrero, Manchado, \& Serra-Ricart 1996) which also exhibits the striking
``spiral arm'' morphology of the halo of NGC~6337).  
>From the symmetry of the velocity field, we derive a heliocentric systemic
velocity $V_{\rm sys}$=$-70$$\pm$$4$~\kms, in excellent agreement with the
measurement of $-71$$\pm$$4$~\kms\ by Meatheringham, Wood, \& Faulkner
(1988).

Filaments $A$ and $B$ have radial velocities of $\sim -40$~\kms\ and $\sim
+60$~\kms, respectively,
 with respect to the systemic velocity.  Their velocities are not
peculiar with respect to the surrounding gas, in the sense that they follow
the overall \ha\ and \nii\ kinematic behavior which characterizes the
emission from the diffuse halo and ring.  At the position of $A$ intersected
by the slit, the \nii\ line broadens to 0.034~nm FWHM (corrected for the
instrumental profile), compared to 0.019~nm in the surrounding gas.
Along $B$, the line profile is also slightly broadened (0.026~nm), or
split into two components separated by $\sim$15~\kms.

It is difficult to determine the real 3-D geometry of the bright ring.  That
it is merely the projection on the sky of a spheroidal shell seems to be
unlikely, since we do not observe the typical elliptical kinematical figure
expected for such a geometry. The \ha\ and \nii\ velocity pattern in the ring
at both position angles, however, is quite complex, and in its innermost
regions both the \ha\ and, to a lesser extent the \nii\ lines, are split into
two components separated by up to 50~\kms\ (see also Fig.~\ref{F-n6337C}).
This clearly suggests that the ring is not thin, but is instead a thick
structure or an ``equatorial bulge'' as observed in several bipolar PNe
(e.g., M~1-28, K~3-46, K~3-72, and NGC~650-1; cf. Corradi \& Schwarz 1995).
The small difference in radial velocity from one side of the ring to the
other indicates that the structure has a low inclination with respect to the
plane of the sky, supporting the hypothesis that the nebula is seen almost
pole on.

Line profiles are also composite through the radial filament $C$
(Fig.~\ref{F-n6337C}), but no peculiar velocities are found as compared to
the ring emission at P.A.=$-39$\gr, where no low-ionization features are
present. Note that the innermost part of $C$ is also slightly visible in 
\oiii, suggesting  a real density enhancement.

\subsection{Discussion: the large-scale structure}

The present data reveal, for the first time, the existence of a high-velocity
($\ge200$\-~\kms) outflow in NGC~6337.  According to the images and spectra,
and considering the similarity with K~4-55 (Guerrero et al. 1996), the most
likely interpretation is that these high-velocity components represent the
polar expansion velocities of a bipolar PN seen almost pole on, the bright
ring being the denser ionized in the equatorial plane (a thick ring or a
``bulge'') from which two symmetrical lobes depart.

The high velocities observed, one order of magnitude larger than those
typical of PNe, are nevertheless not unusual among bipolar objects (Corradi
\& Schwarz 1995).  Also the ``spiral arms'' observed in the \oiii\ image of
NGC~6337 (and in K~4-55) might correspond to the point-symmetrical
distribution of light emission which characterizes the lobes of several
bipolar PNe seen more edge on (e.g, Hb~5, and NGC~6537; Corradi \& Schwarz
1993b). In different nebulae, we would then observe the same overall helical
geometry but from different points of view.  The origin of this ``secondary''
point-symmetrical brightness distribution within the ``primary''
axisymmetrical geometry of the bipolar lobes has rarely been discussed in the
literature.  Pioneering work was done by Pi\c smi\c s, who, 25 years ago,
proposed the occurrence of episodic, highly collimated and magnetized ejecta
from diametrically opposed active spots, in hot stars and PN nuclei (Pi\c
smi\c s 1974; see also Recillas-Cruz \& Pi\c smi\c s 1981). More recently,
the occurrence of collimated, high-velocity and point-symmetrical outflows
was ascribed to the action of fast winds from precessing accretion disks in
binary systems (e.g., Schwarz 1992b).  Cliffe et al.  (1995) presented a
gas-dynamical model in which this kind of morphology is reproduced as the
result of a precessing jet interacting with circumstellar material.  In their
view, the precessing jet would correspond to the brighter point-symmetrical
component of the nebula, while the axisymmetrical bipolar lobes would be the
merging of the bow-shocks of individual jet segments.  Can this scenario be
applied to all bipolar PNe with a point-symmetrical brightness distribution
within lobes?  Considering the sample of bipolar PNe in Corradi \& Schwarz
(1995), and including other recent work on individual objects, it appears
that about 30\% of all bipolar PNe show a more or less pronounced
point-symmetry in the surface brightness of the lobes. Except for special
cases (such as NGC~6309, Schwarz et al.  1992, and perhaps He~2-186, see
Sect.~4), the point-symmetrical light distribution in those bipolar PNe seems
less pronounced than that displayed in figure 2 of Cliffe et al.  (1995).
Further and more complete modeling is clearly needed, and in particular the
resulting velocity field, which might contain characteristic signatures of
the precessing jet model, should be worked out.  This is a very interesting
issue, since should it turn out to be true, we would end with the quite
unexpected conclusion that the overall bi-lobal shape of a significant
fraction of bipolar PNe is the product of the dynamical interaction of
precessing jets ejected by the central stars with the ambient medium.  This
is in contrast with the usual scenario of interacting-winds theories, in
which the bipolar lobes are a result of the aspherical propagation of the
shock driven by an (even isotropic) fast wind starting from a very flattened
AGB mass deposition (cf. Mellema 1997).  Modeling along the lines of the
interacting winds theory is presented by Garcia-Segura (1997, and private
communication), who includes the effects of magnetic fields, stellar
rotation, and precession of the central star induced by wide companions: as a
result, precessing jets and point-symmetric nebular shapes are
formed. Further modeling along this line is desirable.

\subsection{Discussion: the low-ionization features}

In the halo of NGC 6337, there are two peculiar low-ionization filaments,
labeled  $A$ and $B$ in Fig. 1, whose nature is unclear.  As discussed
in the preceding sections, they have velocities similar to those of their
environment, which indicates that they might be the result of in-situ
Rayleigh--Taylor and Kelvin--Helmholtz instabilities occurring in the ``lobes''
of the nebula (see Jones, Kang, \& Tregillis 1994, for a discussion of the
characteristic timescales).  Their position in the kinematic plots,
however, would indicate that they are not located along the polar axis of the
bipolar nebula, but rather at intermediate latitudes on the ``sides'' of the
lobes.

In the inner ring, there are a number of radial filaments, particularly clear
in the \nii/\oiii\ ratio image, the most pronounced one being labeled
$C$. Following the fact that this feature is also slightly visible in \oiii,
it could be a real density enhancement.  The irregular shape of the ring and
the low-ionization radial filaments might then be the result of density
inhomogeneities in the circumstellar density distribution.  The interaction
between the ionization front and the previously formed density fluctuations
can produce the radial tails. The idea, explored by Soker (1998), is that the
ionization front will first form an ionization shadow, which looks like a
tail pointing outward, and as a secondary consequence, a real condensation,
due to the shock compression driven into the tail by the higher pressure of
the surroundings. However, on short time scales the photoionization in the
condensation can smooth out the density inhomogeneity, and therefore these
kinds of structures should be considered as transient features.  An
alternative explanation for the formation of radial filaments was described
by Garc\'\i{a}-Segura \& Franco (1996). It is based on the presence of
dynamical instabilities of radiative shocks, which would lead to the
fragmentation of the shell without involving previous density
inhomogeneities.  In this case, the presence of the ionization front
exacerbates the growth of the instabilities, so that the resulting
filamentary radial structures are not expected to be transient features.  The
model proposed by Dyson et al. (1993) for the low-ionization structures at
small- and intermediate-scales of the Helix nebula (NGC 7293) might also
apply.  According to this model, radially aligned filamentary structures can
be generated as an effect of the interaction of the post-AGB wind, in the
supersonic and subsonic velocity regimes, with dense pre-existing clumps in
the AGB wind. Within this kind of models, short and stubby tails as well as
long and thin ones can be formed, but the details of these processes are still 
to be worked out numerically.

\section{He 2-186}

He~2-186 is a small, poorly studied southern PN. \ha+\nii\ and \oiii\
images are presented in the catalog of Schwarz et al. (1992), which first
revealed the existence of a pair of low-ionization knots detached from the
core of the nebula. A preliminary report on the high-velocity nature of these
knots was given in Schwarz (1993) and Corradi, Schwarz, \& Stanghellini
(1993).  The statistical distance for this object is in the range of 3.5 to
8.2 kpc (Zhang 1995; Acker et al.  1992).

\subsection{Morphology}

\placefigure{F-he2186i}
\placefigure{F-he2186s}

We show in Fig.~\ref{F-he2186i} the \nii\ HST image of He~2-186.  The main
body of the nebula is composed of several bright arcs of emission, which
appear as the limb-brightnened projection of hollow bubbles, with two
``conical'' extensions roughly along the EW direction.  Along P.A.=$+30$\gr\
and detached from the main shell of the nebula, there are two bright,
elongated knots, which are labeled $A$ and $B$ in Fig.~\ref{F-he2186i} and
are partially connected to the core by an S-shaped lane of emission. This
lane is in turn partially resolved into knots. In the ground-based image by
Schwarz et al. (1992), all these features appear to be embedded in a diffuse
nebulosity.  The separation of the knots is of 8$''$.9, and they are also
observed in the higher-ionization species of \oiii\ (see Schwarz et
al. 1992), but are considerably fainter.

\subsection{Kinematics}

The spectrograph slit of NTT was passing both through the center of the
nebula and the knots (P.A.=$-150$\gr), and approximately along the long axis
of the main body of the nebula (P.A.=$-92$\gr), as indicated in
Fig.~\ref{F-he2186i}.  The \nii\ and \ha\ spectra are shown in
Fig.~\ref{F-he2186s}; the \sii\ doublet is not shown, since it is nearly
identical to \nii\ but fainter. \ha\ and \nii\ velocities measured by
Gaussian fitting are also presented in Figure~\ref{F-he2186s}.  From the
symmetry of the position--velocity plot, we adopt a heliocentric systemic
velocity $V_{\rm sys}$=$-81$$\pm$$8$~\kms, in agreement with the value of
$-87$$\pm$$15$~\kms\ measured by Beaulieu (1997), and somewhat higher than
the value of $-67$$\pm$$9$~\kms\ quoted in Schneider et al. (1983) .

Like NGC~6337, He~2-186 shows large expansion velocities.  Knots $A$ and $B$
are redshifted and blueshifted, respectively, by $\sim$135 \- ~\kms\ with respect
to the systemic velocity.  The S-shaped lane of emission which extends toward
the core is partially recorded in the spectrum at P.A.=$-150$\gr, and
exhibits radial velocities which slightly decrease inwards from the values
measured at the knots' positions.  The same spectrum, however, also shows a
peculiar velocity component with reversed redshift/blueshift as compared to
the knots and showing a linear and continuum increase with radius of radial
velocities up to the value of $\pm$$45$~\kms.  This component is not
identified in the HST image, although it can in principle be associated with
the diffuse nebulosity embedding the core and knots which is observed in the
images of Schwarz et al. (1992). Extending out to the the radial distance of
the knots, this second velocity component clearly indicates that the nebula
of He~2-186 has a complex 3-D structure of which we see in projection two
kinematical components.

There is additional spectral features which render the picture even more
complex. The spectrum at P.A.=$-92$\gr\ shows that the \ha\ and \nii\
emission extend much farther than expected from the image, out to at least
10$''$ from the center. In addition, in that spectrum there are a pair of
opposite symmetrical features at $\sim$$\pm$$130$~\kms\ and about 1$''$.9
from the center. Due to the lower spatial resolution of the NTT spectra
(about 1$''$.4 along the direction of the 1$''$ wide slit), and considering
seeing and tracking effects, it is difficult to identify these velocity
components with features in the HST image.

\subsection{Discussion}

If we consider the knots and their S-shaped connection to the core by
themselves, the present data would suggest that the knots of He~2-186 are at
the leading tip of an extended, bent, highly collimated and high-velocity
outflow. This is indicated by the S-shaped morphology of the knot--core
connection and of its velocity pattern. The most likely cause of the bending
would be precession of the collimating source. Again, models like that of
Cliffe et al. (1995) or Garc\'\i a-Segura (1997) might apply in this case.

Spectroscopy, however, unveils other components, and renders the situation
more difficult to fit into the above scenario. The additional velocity
components extending as far as the knots but with reversed redshift clearly
indicate that the extended outflow of He~2-186 has a complex 3-D shape.  The
gross morphology of the velocity field is somewhat reminiscent of that of the
``Southern Crab'', the nebula around the symbiotic star He~2-104 (Corradi \&
Schwarz 1993a), which possesses a bipolar nebula with prominent polar jets
(cf. Fig.~3 in Corradi \& Schwarz 1993a and Fig.~\ref{F-he2186s} in this
paper).  It might be that He~2-186 is a similar object poorly
resolved. Alternatively, as discussed for NGC~6337 in Sect.~3.3, the knots
and lanes of He~2-186 might be prominent point-symmetrical features within an
overall structure consisting of (faint) hollow bipolar lobes.

\section{K 4-47}

K 4-47 is an almost unstudied northern PN. No previous individual studies
exist in the literature apart from measurements in radio or infrared
surveys. This PN is associated with the {\it IRAS} source 04166+5611 and
shows 5-GHz radio emission (Aaquist \& Kwok 1990).  Cahn, Kaler, \&
Stanghellini (1992) give a statistical distance of 8.5 kpc.  Using another
statistical method, Zhang (1995) obtained a distance of more than 20~kpc (and
van de Steene \& Zijlstra (1994) give an upper limit of 26 kpc); these are
unrealistic values putting the nebula well outside the boundaries of the
Galaxy (see the discussion in the following sections).

\subsection{Morphology}

\placefigure{F-k447i}

The narrow-band images in Fig.~\ref{F-k447i} reveal that K~4-47 consists of
an emission-line core and two diametrically
opposite blobs aligned at P.A.=$+41$\gr.
The blobs are prominent in the low-ionization lines of \nii\ and \sii, visible in
\ha, and absent in \oiii, while the core is of relatively higher excitation,
showing  bright \oiii\ emission.  A faint lane of emission connects the
blobs to the core.  The blobs and core are practically unresolved in our image,
and have a FWHM only slightly larger than the seeing value (0.$''$8).  
The blobs
are not symmetrically located with respect to the core, the northern one
being at larger projected distance than the southern one (see Table~2).  
Continuum emission is detected neither in the core nor in the knots in our \ha\
off-band image (not presented here).

Our images also suggest the existence of other small differences in the
spatial location of the core and the blobs, such as a displacement of 0$''$.3
southwest of the Gaussian centroid of the core in \nii\ with respect to
\ha.  The \nii\ center of the southern blob also seems to lie about 0$''$.2
more southwest than in \ha\ and \sii, but these latter findings should be
confirmed by new imaging with better resolution and high astrometric
quality.

\subsection{Kinematics}

The echelle spectrum at P.A.=$+41$\gr\ is \- shown in Fig.~\ref{F-k447s},
together with the heliocentric \ha\ and \nii\ velocities measured by Gaussian
fitting along the slit. Radial velocities and line widths for the blobs are
also reported in Table~2.

The gross kinematic figure is that of a linear increase of radial
velocities from the core to the blobs, but several peculiar kinematic
features are also observed, especially in  \ha.  The velocity
separation of the two blobs is  105~\kms\ in \ha\, and  115~\kms\ in
\nii\ (note that this difference is larger than the measurement errors).
Line profiles at the positions of the blobs are roughly Gaussian 
and are significantly broadened, with a FWHM, corrected for the instrumental
profile, of 75-90~\kms (see Table~2).

The \ha\ radial velocities are systematically redshifted around the core
position as compared to the \nii\ ones, by up to 30~\kms. This difference is
noticeable, and together with the displacement of the \ha\ peak position from
the \nii\ one, seen in the images, suggests the existence of two distinct
flows of matter with different geometries and/or velocities, one bright in
\ha\ and the other with enhanced \nii\ emission. In this respect, note also
the peculiar kinematic feature, detected in \ha\ 1$''$.5 southwest of the
core, which has a radial velocity of $\sim +70$~\kms\ with respect to the
central region and shows sign reversal with respect to the collimated outflow
of the blob on the same side of the nebula.

The complex kinematics in the innermost regions also makes it difficult to
estimate the systemic velocity of the nebula, and compute the expansion
velocity of each blob separately.  If we take the average radial velocity of
the blobs as the heliocentric systemic velocity of K~4-47, we obtain a value
of $-32$~\kms. If we instead assume that the different distances from the
center of each blob just reflect different initial velocities of coeval
ejecta, the requirement of proportionality between distance and velocity
would imply a systemic velocity of $-42$~\kms. We adopt a mean value, 
$V_{\rm sys}$=$-37$$\pm$$10$~\kms, as the heliocentric systemic velocity of
K~4-47, the large error reflecting the above uncertainties.

According to both imaging and spectros\-co\-py, the separation of the two blobs
at peak emission of Gaussian fitting is of 7$''$.5$\pm$0$''$.1 in the bright 
\nii\ line.

\placefigure{F-k447s}

\subsection{Discussion} 

K~4-47 is composed of a bright emission-line core exhibiting complex
kinematics. Departing from the core, a highly collimated outflow ending in
two low-ionization blobs is detected.  The velocity separation of the 
blobs projected onto the line of sight is around 110~\kms, and their \nii\ and
\ha\ line profiles are substantially broadened (75 -- \- 100~\kms\ FWHM).

The blobs might be the tips of jets interacting with the ambient medium, or
bullets of dense gas ejected from the central star and expanding at
considerable velocities. If so, radiating bow-shocks are expected to
form. Hartigan et al. (1987) constructed line profiles for
shocked--ionized bullets, whose line widths are shown to be a direct measure
of the shock velocity. Using their prescriptions, and correcting the observed
line widths for instrumental and thermal broadening (assuming 
$T_{\rm e}$=10000~K
and zero turbulent velocity), we obtain shock velocities between 125 and
155~\kms, depending on the blob and the line (\ha\ or \nii)
considered. Assuming that blobs are expanding through a stationary
circumstellar medium (the interstellar medium or the slowly expanding remnant
of the red-giant wind), the shock velocity is nothing but the expansion
velocity of the blobs, and from the observed projected velocities we derive
an inclination of the outflow to the line of sight of 65\gr -- 70\gr.  According
to Hartigan et al. (1987), such a mild inclination to the plane of the sky
and the computed shock velocity around 150~\kms\ would yield quite symmetric
integrated line profiles, as is indeed observed in the spectra of the blobs of
K~4-47. Note however that the main difference from the models by Hartigan et
al. (1987) is that in the case of K~4-47 the ionization balance in the blobs
might be dominated by the energetic radiation from the central star, whereas 
their models are for purely shocked--ionized bullets such as H--H objects.

The distance to K~4-47 is  unknown, and considering the peculiar
morphology of the nebula, standard statistical methods are likely to give
completely unreliable results.  We could estimate rough limits on the
distance to K~4-47 by assuming that it participates in the general circular
rotation around the Galactic center.  With a standard Galactic rotation curve
and with the adopted systemic velocity of $-37$$\pm$$10$~\kms, the
kinematical distance of K~4-47 is constrained between 3 and 7~kpc.  Note that
a distance larger than 7~kpc is unlikely, since K~4-47 is located nearly in
the direction of the Galactic anticenter ($l$=149\gr), and larger distances
would position the nebula in very peripheral regions of the Galaxy (the
truncation radius of the volume density of stars in the Galaxy is often
assumed to be around 15~kpc, cf. Wainscoat et al. 1992).

With these distance limits, and using the orientation and kinematic figures
above, the linear size of K~4-47 would be  $\sim$0.1~pc for a distance of
3~kpc, and  $\sim$0.3~pc for 7~kpc. The age of the blobs would be
$\sim$400~yr or $\sim$900~yr, for the short and long distances, respectively.

K~4-47 shares several similarities with the PN M~1-16 (Schwarz 1992a, Corradi
\& Schwarz 1993c; Aspin et al. 1993), in which a shocked, highly collimated
and high-velocity outflow was also detected. In M~1-16, the outflow is
resolved into a pair of bipolar lobes ending in a series of relatively bright
knots. Imaging with {\it HST} resolution would be needed to resolve similar
structures in K~4-47 if they existed.

\section{Conclusions}

A morphological and kinematical study of the PNe NGC~6337, He~2-186, and
K~4-47 is presented. All three nebulae show gas expanding at large velocities
(130--200 \kms). These high-velocity outflows define the main collimation
axis of the ejecta.

In the case of He~2-186 and K~4-47, the high-velocity outflows correspond to
a pair of opposed symmetrical, low-ionization knots detached from the cores
of the nebulae. Evidence of strong shocks is found in the knots of K~4-47,
which might be one of the causes of the enhancement of the emission from
low-ionization species. Low-resolution spectra, as well as detailed modeling
taking into account all the different mechanism which play a role in the
production of spectral lines (abundances, physical properties, shocks, and
photoionization from the central star) are clearly needed.  In any case, the
mass-loss scenario which is derived from the present observations is quite
peculiar, especially for K~4-47. In this object, and in similar ones (e.g.,
M~1-16, Schwarz 1992a), all the ionized mass appears to be contained in the
compact core and in the high-velocity blobs.  Fast-moving bullets and jets
have been observed also in other PNe (e.g., NGC~3918, Corradi et al.  1999),
but generally they are secondary morphological structures as compared to the
main shells of the nebulae. In K~4-47, on the contrary, the core--jets--blobs
constitute the whole nebula. Is the {\it entire} AGB envelope constrained to
flow within such a small solid angle? The interacting-winds models of Icke et
al. (1992) and Garc\'\i{a}-Segura et al. (1999) are able to explain the
formation of highly collimated nebulae, starting from torus-like initial
density distributions or in the presence of significant magnetic fields, but
it is not clear whether they can reproduce extreme morphologies such as those
of K~4-47 and M~1-16.  An alternative hypothesis, is that objects like K~4-47
are not genuine PNe, but are the results of (sporadic?) mass loss from
interacting binary systems.  There, accretion disks are expected to provide
the conditions for extreme collimation of the outflows. In erupting systems,
such as symbiotic stars (see Corradi et al. 1999 for a discussion of the
properties of the outflows from this class of object), high velocity winds
from the accreting components are also expected.  Thus it will be important
to investigate in depth the nature of the central star of K~4-47 and similar
objects, searching for possible signatures of binarity.

In the case of NGC~6337, the highest velocities measured ($200$~\kms), which
are thought to occur in the polar regions of bipolar lobes seen almost pole
on, are not associated with the outer low-ionization filaments observed in
\nii. The location, shape and velocities of these outer filaments pose a
difficult problem in order to explain their origin. The low-ionization knots
and tails observed in the bright ``equatorial'' ring are instead interpreted
as being the result of initial dynamical instabilities which lead to
fragmentation of the ring and which are enhanced by the ionization front
(according to the modeling of Garc\'\i a-Segura \& Franco 1996), or, on the
other hand, transient density enhancments due to the interaction of the
ionization front with previous density fluctuations in the ISM (Soker 1998).

Finally, both the point-symmetrical morphologies of NGC~6337 and He~2-186
streng\-then the idea of the occurrence of precessing outflows in PNe, a
hypothesis that has found more and more observational support in the recent
years (e.g., Schwarz 1992a; Lopez, \- Meaburn \& Palmer 1993; Guerrero \&
Manchado 1998), and which also naturally leads to the idea of the existence
of binary systems as the central stars of these nebulae.  Embarrasingly, in
spite of the numerous high-quality data which are presently available it
remains mysterious why these point-symmetrical features are often found
within bipolar lobes with an overall axisymmetrical geometry. Cle\-arly, some
basic piece of the puzzle of the PN formation and evolution is still
missing.

\clearpage

\section{Acknowledgments}

We thank Benjamin Montesinos for giving us the observing time at the
NOT+IACUB.  The work of RLMC, EV, and AM is supported by a grant of the
Spanish DGES PB97-1435-C02-01, and that of DRG by a grant from the Brazilian
Agency Funda\c c\~ao de Amparo \`a Pesquisa do Estado de S\~ao Paulo 
(FAPESP; 98/07502-0).

\clearpage
\figcaption[ ]{The NTT images of NGC~6337, on a logarithmic intensity scale.
Above: the \oiii\ image, at different intensity levels to highlight both the  
inner ring of the nebula (left) and its faint `spiral-shaped'  halo (right).
Below: the \nii\ image and the \nii/\oiii\ ratio map.  The locations
of the slit used for spectroscopy are indicated in the \nii\ image by short
lines on either side of the object.
\label{F-n6337i}}
\figcaption[ ]{
    Left: the NTT long-slit spectra of NGC~6337, on a logarithmic
    intensity scale. The faint emission to the left of \ha\ is an \heii\ line
    at $\lambda$=656.01~nm. Right: radial heliocentric velocities
    computed by Gaussian fitting of the \ha\ (empty circles) and \nii\ (full
    circles) lines. The dotted vertical line is the adopted systemic
    velocity.
\label{F-n6337s}}
\figcaption[ ]{
    Details of the images (the two leftmost boxes) and of the spectrum at
    P.A.=$-75$\gr\ (the three rightmost boxes) of NGC~6337 for the region
    around the low-ionization feature $C$.  The images have been rotated so
    as to have the slit location along the vertical direction and to allow
    for direct comparison with the spectra.  Velocities are as in
    Figure~\ref{F-n6337s}.
\label{F-n6337C}}
\figcaption[ ]{The \nii\ HST image of He~2-186, on a logarithmic intensity
    scale, and with different intensity cuts for the inner and outer regions
    of the nebula. The locations of the slit are indicated by lines on either
    side of the object. We also mark distances of 5$''$ from the center along
    the slit directions, to facilitate comparison with the spectra in
    Fig.~5.
\label{F-he2186i}}
\figcaption[ ]{
    Left: the spectra of He~2-186 on a logarithmic scale. 
    Right: measured \ha\ (empty circles) and \nii\ (full
    circles) radial velocities, corrected for the adopted systemic velocity
    of the nebula.
\label{F-he2186s}}
\figcaption[ ]{
    The NOT images of K~4-47 on a linear scale.
\label{F-k447i}}
\figcaption[ ]{
    Left: the spectrum of K~4-47 at P.A.=$+41$\gr\ on a linear
    scale. Right: heliocentric \ha\ (empty circles) and \nii\
   (full circles) radial velocities.}
\label{F-k447s}

\newpage
\begin{deluxetable}{lclc}
\tablenum{1}
\tablewidth{40pc}
\tablecaption{Log of the observations}
\tablehead{
\multicolumn{4}{c}{\bf\it Images} \\
\multicolumn{1}{l}{Object}& 
\multicolumn{1}{l}{Telescope} & 
\multicolumn{1}{l}{Filter (exposure time, min)} &
\multicolumn{1}{c}{Seeing}}

\startdata
NGC 6337  & NTT & \nii\ (4), \oiii\ (2)    & 0$''$.9  \nl
K 4--47   & NOT & \nii\ (10), \ha\ (10), \ha\ cont. (5), \sii\ (10) 
& 0$''$.8 \nl
          &     & \oiii\ (10) & 0$''$.9 \nl
&  & & \nl
\multicolumn{4}{c}{\bf\it Long--slit spectra} \nl
       &           & P.A. (exposure time, min) &        \nl
\hline \nl
NGC 6337  & NTT  &  $-39$\gr\ (60), $-75$\gr\ (30)   & 0$''$.9 \nl
He 2-186 & NTT  &  $-150$\gr\ (5), $-92$\gr\ (5)    & 0$''$.9 \nl
K 4-47   & NOT  &  $41$\gr\ (60)                     & $\sim$1$''$ \nl
\enddata
\end{deluxetable}

\newpage 
\begin{deluxetable}{lccccccc}
\tablenum{2}
\tablewidth{24pc}
\tablecaption{Distances, velocities and line widths for the knots of K 4-47.}
\tablehead{
\multicolumn{1}{l}{} &
\multicolumn{2}{c}{d [$''$]}& 
\multicolumn{2}{c}{V$_r$ [\kms]} & 
\multicolumn{2}{c}{FWHM [\kms]} \nl
\multicolumn{1}{l}{} &
\multicolumn{1}{c}{\ha} &
\multicolumn{1}{c}{\nii} &
\multicolumn{1}{c}{\ha} &
\multicolumn{1}{c}{\nii} &
\multicolumn{1}{c}{\ha} &
\multicolumn{1}{c}{\nii}
} 

\startdata
South         & $3.1$ & $3.0$\tablenotemark{a} &$-86$ & $-89$ & $88$ & $75$ \nl
North         & $4.2$ & $4.5$\tablenotemark{a} & $20$ & $28$  & $99$ & $82$ \nl
\enddata
\tablenotetext{a}{From both imaging and spectroscopy.}
\tablecomments{$d$ is the apparent distance measured in the image 
from the peak emission of the core, V$_r$ 
the measured heliocentric velocity, and FWHM the full width at half maximum
of line emission corrected for the instrumental profile.}
\end{deluxetable}

\end{document}